\documentclass[12pt,letterpaper]{article}
\usepackage[usenames, dvipsnames]{xcolor}
\usepackage{jcapmod}

\usepackage{tocloft}
\usepackage[]{todonotes}
\usepackage{verbatim}
\usepackage{mathrsfs}
\usepackage{cleveref}
\usepackage[shortlabels]{enumitem}
\usepackage{pgf}
\usepackage{tikz-cd}
\usetikzlibrary{shapes,arrows}
\usetikzlibrary{calc}
\usepackage{pgfplots}
\pgfplotsset{compat=1.12}
\usepackage{tikz-3dplot}

\usepackage{tikz}
\usepackage{setspace,caption}
\usepackage{lipsum}
\usetikzlibrary{matrix,arrows,calc}
\makeatletter
\newsavebox\myboxA
\newsavebox\myboxB
\newlength\mylenA

\definecolor{cornellRed}{HTML}{B31B1B}
\definecolor{cornellBlue}{HTML}{0068AC}
\definecolor{cornellGreen}{HTML}{6EB43F}

\newtheorem{proposition}{Proposition}

\usepackage{bbm} 					
\usepackage{slashed} 				
\usepackage{graphicx}				
\usepackage{subcaption}			
\usepackage{psfrag}				
\usepackage{tensor}				
\usepackage{fouridx}				
\usepackage{bm}					
\usepackage{mdframed}				
\usepackage{multirow}				
\usepackage{soul}					
\usepackage{bbold}				
\usepackage{multicol}				
\usepackage{tikz-cd}
\usepackage{rotating}

\usetikzlibrary{arrows}

\tikzset{
commutative diagrams/.cd,
arrow style=tikz,
diagrams={>=latex}}

\usepackage{amsmath}
\usepackage{amssymb}
\usepackage{amsfonts}
\usepackage{mathtools}

\usepackage{feynmf}
\usepackage{marvosym}

\usepackage{import}

\newcommand*\xoverline[2][0.75]{%
    \sbox{\myboxA}{$\m@th#2$}%
    \setbox\myboxB\null
    \ht\myboxB=\ht\myboxA%
    \dp\myboxB=\dp\myboxA%
    \wd\myboxB=#1\wd\myboxA
    \sbox\myboxB{$\m@th\overline{\copy\myboxB}$}
    \setlength\mylenA{\the\wd\myboxA}
    \addtolength\mylenA{-\the\wd\myboxB}%
    \ifdim\wd\myboxB<\wd\myboxA%
       \rlap{\hskip 0.5\mylenA\usebox\myboxB}{\usebox\myboxA}%
    \else
        \hskip -0.5\mylenA\rlap{\usebox\myboxA}{\hskip 0.5\mylenA\usebox\myboxB}%
    \fi}
\makeatother

\definecolor{cobalt}{RGB}{44, 98, 120}
\definecolor{celadon}{rgb}{0.67, 0.88, 0.69}
\definecolor{dm}{cmyk}{.20, 0, .30, 0}
\definecolor{burgundy}{rgb}{0.5, 0.0, 0.13}
\definecolor{plotBlue}{RGB}{94, 130, 181}

\hypersetup{
  colorlinks,
  citecolor=Violet,
  linkcolor=cobalt,
  urlcolor=Blue}

\DeclareSymbolFontAlphabet{\mathbb}{AMSb}

\newif\iffastcompile

\fastcompilefalse

\iffastcompile
\newcommand{\mk}[1]{}
\newcommand{\lm}[1]{}
\else
\newcommand{\mk}[1]{\todo[color=burgundy!30, size=\scriptsize, bordercolor=burgundy!30]{MK: #1}}
\newcommand{\lm}[1]{\todo[color=dm!90, size=\scriptsize, bordercolor=dm!90]{LM: #1}}

\fi

\ProvideTextCommandDefault{\Dbar}{%
\leavevmode\lower.5ex\rlap{\hskip-.07em\accent"16}D%
}

\begin{document}
	\newcommand{\main}{.}
\begin{titlepage}

\setcounter{page}{1} \baselineskip=15.5pt \thispagestyle{empty}
\setcounter{tocdepth}{2}

\bigskip\

\vspace{1cm}
\begin{center}
{\fontsize{18}{24} \bfseries A Note on $h^{2,1}$ of Divisors in CY Fourfolds I}
\end{center}

\vspace{0.55cm}

\begin{center}
\scalebox{0.95}[0.95]{{\fontsize{14}{30}\selectfont Manki Kim$^{a,b}$\vspace{0.25cm}}}

\end{center}

\begin{center}

\vspace{0.15 cm}
{\fontsize{11}{30}
\textsl{$^{a}$Department of Physics, Cornell University, Ithaca, NY 14853}\\
\textsl{$^{b}$Center for Theoretical Physics, MIT, Cambridge, MA 02139}}\\
\vspace{0.25cm}

\vskip .5cm
\end{center}

\vspace{0.8cm}
\noindent

In this note, we prove combinatorial formulas for $h^{2,1}$ of prime toric divisors in an arbitrary toric hypersurface Calabi-Yau fourfold $Y_4.$ We show that it is possible to find a toric hypersurface Calabi-Yau in which there are more than $h^{1,1}(Y_4)$ non-perturbative superpotential terms with trivial intermediate Jacobian. Hodge numbers of divisors in toric CICYs are the subjects of the part two.

\vspace{1.1cm}

\vspace{3.1cm}

\noindent\today

\end{titlepage}
\tableofcontents\newpage

\section{Introduction}
The nature of dark energy in quantum gravity remains a great mystery. In order to understand dark energy in string theory, one may attempt to find de Sitter solutions. Kachru, Kallosh, Linde, and Trivedi (KKLT) argued that one can find meta-stable de Sitter solutions in type IIB string theory if the right modules are in place in a given string compactification \cite{Kachru:2003aw}.\footnote{For a similar proposal see \cite{Balasubramanian:2005zx}.}


Over the last few years, we have gathered significant evidence that light complex structure moduli in KKLT like solutions are rather more generic than it was assumed before \cite{Bena:2018fqc,Randall:2019ent,Dudas:2019pls,Demirtas:2019sip,Demirtas:2020ffz,Blumenhagen:2020ire}. In light of this developement, it may no longer qualify as a good practice to assume that the one-loop pfaffian of the non-perturbative superpotential \cite{Witten:1996bn,Witten:1996hc} is constant without an explicit demonstration that the one-loop pfaffian is well controlled, \emph{even} in the absence of the open string moduli.

In this work, we are interested in the one-loop pfaffian of the non-perturbative superpotentials. Although the type IIB description and the F-theory description of the non-perturbative superpotentials are physically equivalent, we will mostly focus on the F-theory description \cite{Vafa:1996xn} because F-theory provides a clean geometric picture.

In M/F-theory, the non-perturbative superpotentials are genated by M5-branes wrapped on rigid divisors. The partition function of an M5-brane on a divisor $D$ is known to be a section of a line bundle $\mathcal{L}$ whose Chern class is the principal polarization of the intermediate Jacobian $\mathcal{J}_D:=H^3(D,\Bbb{R})/H^3(D,\Bbb{Z})$ \cite{Witten:1996hc,Belov:2006jd}. The first step in determining the one-loop pfaffian of the non-perturbative superpotential is therefore to compute the intermediate Jacobian $\mathcal{J}_D.$

In order to compute the intermediate Jacobian $\mathcal{J}_D,$ one must compute its dimension and complex structure moduli. To compute the dimension of $\mathcal{J}_D,$ one may compute the middle cohomology via sequence chasing \cite{Denef:2005mm}. Albeit it is possible to compute the dimension of $\mathcal{J}_D$ in this way \emph{in principle}, this method becomes computationally intractable for Calabi-Yau manifolds with large hodge numbers. Therefore, it is preferable to have combinatorial formulas for the dimension of $\mathcal{J}_D$ to carry out an extensive search for Calabi-Yau manifolds with desirable properties.

In this work, we prove combinatorial formulas to determine the dimension of $\mathcal{J}_D$ for prime toric divisors in toric hypersurface Calabi-Yau 4-folds, which we will denote by $Y_4.$\footnote{For an interesting series of works on the period in the context of tropical geometry, see \cite{yamamoto2018periods,yamamoto2021tropical,ruddat2021homology,ruddat2020period}.} Then, we construct a new class of F-theory compactifications. We will present a more detailed study of such F-theory uplifts in \cite{ftheory}. We furthermore show that in relatively simple F-theory compactifcations, there are more than $h^{1,1}(Y_4)$ non-perturbative superpotential terms with trivial intermediate Jacobian. 
We will address the Freed-Witten anomaly \cite{Witten:1996md,Freed:1999vc} in \cite{ftheory}. The hodge numbers of divisors in toric CICYs will be studied in \cite{cicy}.

The organization of this paper is as follows. In \S\ref{S:results}, we summarize the combinatorial formulas for $h^{2,1}$ and $h^\bullet(D,\mathcal{O}_D)$ of prime toric divisors. In \S\ref{S:proofs}, we prove the combinatorial formulas. In \S\ref{S:examples}, we study a few examples and show that more than $h^{1,1}(Y_4)$ prime toric divisors support non-perturbative superpotentials. In \S\ref{S:review}, we review relevant materials. Readers familiar with stratifications and toric geometries may skip this appendix.

\section{Summary}\label{S:results}
Consider a pair of reflexive polytopes $(\Delta,\Delta^\circ),$ where $\Delta\in M$ and $\Delta^\circ\in N.$ Let $\mathcal{T}$ be a Fine Regular Star Triangulation (FRST) of $\Delta^\circ.$ We treat $\Delta$ as the Newton polytope for the anti-canonical line bundle of the toric variety $\Bbb{P}_{\Sigma}$ defined by the fan $\Sigma$ over $\mathcal{T}.$ By $t_i,$ we denote $i$-dimensional simplices in $\mathcal{T}\cap \Delta^\circ.$ To each point $p\in\Delta^\circ\cap N,$ we associate a prime toric divisor $D_p$ in the toric hypersurface Calabi-Yau manifold. We define $l^i(\Theta)$ for $\Theta\in\Delta$ to be the number of points in $i$-dimensional simplicies in $\Theta.$ Similarly, by $l^*(\Theta)$ we denote the number of points interior to $\Theta.$ We define genus of a face $\Theta$ to be $g(\Theta):= l^*(\Theta^\circ).$

Let $v$ be a vertex of $\Delta^\circ.$ Similarly, let $v_e,~v_f,~v_s,$ be points interior to an edge $e,$ a 2-face $f,$ and a 3-face $s,$ respectively. Then, we compute
\begin{equation}
-h^{2,1}(D_v)=e^{2,1}(D_v)=l^3(v^\circ)-l^2(v^\circ)-\varphi_2(v^\circ)-\sum_{f\geq v}l^*(f^\circ)\#(f\supset t_1\supset v)\,,\label{res1}
\end{equation}
\begin{equation}
-h^{2,1}(D_{v_e})=e^{2,1}(D_{v_e})=\sum_{f\geq e}l^*(f^\circ)(1-\#(f\supset t_1\supset\nu))\,,\label{res2}
\end{equation}
\begin{equation}
-h^{2,1}(D_{v_f})=e^{2,1}(D_{v_f})=l^*(f^\circ)(2-\#(f\supset t_1\supset \nu ))\,,\label{res3}
\end{equation}
\begin{equation}
-h^{2,1}(D_{v_s})=e^{2,1}(D_{v_s})=0\,,\label{res4}
\end{equation}
where $\#(f\supset t_1\supset v)$ denotes the number of 1-simplices in $f\cap\mathcal{T}$ that contain the point $v,$ and $\varphi_2(v^\circ)=l^*(2v^\circ)-5l^*(v^\circ).$ 
For the hodge vectors, one obtains \cite{Klemm:1996ts,Braun:2016igl,Braun:2017nhi}
\begin{equation}
h^\bullet(D_v,\mathcal{O}(D_v))=(1,0,0,l^*(v^\circ))\,,
\end{equation}
\begin{equation}
h^\bullet(D_{v_e},\mathcal{O}(D_{v_e}))=(1,0,l^*(e^\circ),0)\,,
\end{equation}
\begin{equation}
h^\bullet(D_{v_f},\mathcal{O}(D_{v_f}))=(1,l^*(f^\circ),0,0)\,,
\end{equation}
\begin{equation}
h^\bullet(D_{v_s},\mathcal{O}(D_{v_s}))=(1+l^*(s^\circ),0,0,0)\,.
\end{equation}
We finally remark that one can also compute the combinatorial formulas for $h^{1,1}$ from the stratifications studied in \S\ref{S:proofs}.

\section{Proofs}\label{S:proofs}
In this section, we prove \eqref{res1}-\eqref{res4} by constructing the stratas of prime toric divisors. Let us recall the definitions. By $t_i,$ we denote $i$-dimensional simplices in $\mathcal{T}\cap \Delta^\circ,$ where $\mathcal{F}$ is an FRST of $\Delta^\circ.$ Let $v$ be a vertex of $\Delta^\circ.$ Similarly, we denote an edge, a 2-face, and a 3-face by $e,$ $f,$ and $s,$ respectively.
\subsection{Vertex}
The stratification for the prime toric divisor $\{z_v=0\}\cap \overline{Z}_{\mathcal{F},\Sigma}$ is
\begin{align}
D_v=&Z_{v^\circ}\coprod_{e\geq v} Z_{e^\circ}\times (pt)\coprod_{f\geq v} Z_{f^\circ}\times \left(\sum_{f\supset t_1 \supset v}(C^*)\coprod \sum_{f\supset t_2 \supset v}(pt)\right)\nonumber\\&\coprod_{s\geq v} Z_{s^\circ}\times\left( \sum_{s\supset t_1 \supset v}(C^*)^2\coprod \sum_{s\supset t_2 \supset v} (C^*) \coprod \sum_{s\supset t_3\supset v} (pt)\right).
\end{align}
Because $e^{2,1}(Z_{e^\circ})=e^{2,1}(Z_{f^\circ})=e^{2,1}(Z_{s^\circ})=e^{1,0}(Z_{s^\circ})=0,$ there can be contributions to $e^{2,1}(D_v)$ only from $e^{2,1}(Z_{v^\circ})$ and $e^{1,0}(Z_{f^\circ})\times e^{1,1}(C^*).$ As a result, we obtain
\begin{align}
e^{2,1}(D_v)=&e^{2,1}(Z_{v^\circ})+\sum_{f\geq v}e^{1,0}(Z_{f^\circ})\times \left(\sum_{f\supset t_1\supset v} e^{1,1}(C^*)\right)\\
=&l^3(v^\circ)-l^2(v^\circ)-\varphi_2(v^\circ)-\sum_{f\geq v}l^*(f^\circ)\#(f\supset t_1\supset v).
\end{align}
One can similarly compute the hodge vector \cite{Klemm:1996ts,Braun:2016igl,Braun:2017nhi}
\begin{equation}
h^\bullet(D_v,\mathcal{O}_{D_v})=(1,0,0,l^*(v^\circ)).
\end{equation}

Before we move on to the edge case, let us study a special case: $l^*(v^\circ)=1.$ To simplify the discussion, we will further assume that one of the anti-canonical sections is a square of the toric cooridnate $v.$ When $l^*(v^\circ)=1$ the corresponding divisor is a CY 3-fold. We first recall a useful formula
\begin{equation}
l^*(2\Theta)=l(\Theta),\label{eqn:doubling}
\end{equation}
when $l^*(\Theta)=1.$\footnote{See \cite{Braun:2014xka} for more general treatment.} We thus write
\begin{equation}
h^{2,1}(D_v)=l(v^\circ)-5 -\sum_{\Gamma \leq v^\circ}l^*(\Gamma)+\sum_{F\leq v^\circ} l^*(F)\#(F^\circ \supset t_1\supset v),
\end{equation}
where $\Gamma$ and $F$ are 3-faces and 2-faces in $v^\circ$ respectively. To rewrite $\#(F^\circ\supset t_1\supset v)$ in a more familiar form, let us recall that there is a hyperplane $H$ such that $H\cap\Delta_{5d}^\circ=\Delta_{4d}^\circ.$ Note that there is an isomorphism $\Delta_{4d}\equiv v^\circ.$ Then, it is straightforward to show that $\#(F^\circ \supset t_1\supset v)= l^*(F^\circ \cap H),$ which equals to $l^*(F^\circ\cap \Delta_{4d}^\circ).$  As a result, we reproduce the famous formula for $h^{2,1}$ of a toric hypersurface CY 3-fold!
\subsection{Edge} 
Let $\nu$ be an interior point of an edge $e.$ The stratification for the divisor $D_\nu$ is then
\begin{align}
D_\nu=& Z_{e^\circ}\times (C^*\coprod 2 pts )\coprod_{f\geq e} Z_{f^\circ}\times \left(\sum_{f\supset t_1 \supset \nu}(C^*)\coprod \sum_{f\supset t_2 \supset \nu}(pt)\right)\nonumber\\&\coprod_{s\geq e} Z_{s^\circ}\times\left( \sum_{s\supset t_1 \supset \nu}(C^*)^2\coprod \sum_{s\supset t_2 \supset \nu} (C^*) \coprod \sum_{s\supset t_3\supset \nu} (pt)\right).
\end{align}
For this type of divisors, we obtain contributions to $e^{2,1}(D_\nu)$ only from $Z_{e^\circ}$ and $Z_{f^\circ}$ 
\begin{align}
e^{2,1}(D_\nu)=& e^{1,0}(Z_{e^\circ})e^{1,1}(C^*)+\sum_{f\geq e}e^{1,0}(Z_{f^\circ})\times \left(\sum_{f\supset t_1\supset \nu} e^{1,1}(C^*)\right)\\
=& l^2(e^\circ)-l^1(e^\circ)-\sum_{f\geq e} l^*(f^\circ)\# (f\supset t_1\supset \nu)\\
=& \sum_{f\geq e}l^*(f^\circ)(1-\#(f\supset t_1\supset \nu)).
\end{align}
For the hodge vector, one obtains
\begin{equation}
h^\bullet(D_\nu,\mathcal{O}_{D_\nu})=(1,0,l^*(e^\circ),0).
\end{equation}

\subsection{2-Face}
Let $\nu$ be an interior point of a 2-face $f.$
\begin{align}
D_\nu= & Z_{f^\circ}\times \left((C^*)^2\coprod\sum_{f\supset t_1 \supset \nu}(C^*)\coprod \sum_{f\supset t_2 \supset \nu}(pt)\right)\nonumber\\& \coprod_{s\geq e}Z_{s^\circ}\times\left( \sum_{s\supset t_1 \supset \nu}(C^*)^2\coprod \sum_{s\supset t_2 \supset \nu} (C^*) \coprod \sum_{s\supset t_3\supset \nu} (pt)\right).\label{eqn:2-face stratification}
\end{align}
Similarly, we obtain
\begin{equation}
e^{2,1}(D_\nu)=l^*(f^\circ)(2-\#(f\supset t_1\supset\nu))
\end{equation}
and
\begin{equation}
h^\bullet(D_v,\mathcal{O}_{D_v})=(1,l^*(f^\circ),0,0).
\end{equation}
\subsection{3-Face}
Let $\nu$ be an interior point of a 3-face $s.$
\begin{equation}
D_\nu= Z_{s^\circ}\times\left( (C^*)^3\coprod\sum_{s\supset t_1 \supset \nu}(C^*)^2\coprod \sum_{s\supset t_2 \supset \nu} (C^*) \coprod \sum_{s\supset t_3\supset \nu} (pt)\right).
\end{equation}
From the stratification, it is manifest that $D_\nu$ is a $(1+l^*(s^\circ))$ copies of a toric threefold. Hence, we obtain $e^{2,1}(D_\nu)=0$ and
\begin{equation}
h^\bullet(D_v,\mathcal{O}_{D_v})=(1+l^*(s^\circ),0,0,0).
\end{equation}

\section{Examples}\label{S:examples}
\subsection{The Sextet and the mirror of the Sextet}
We first study the simplest example, the Sextet fourfold. Let $(\Delta,\Delta^\circ)$ be the dual pair of the reflexive polytopes. $\Delta^\circ$ has vertices
\begin{equation}
\Delta^\circ\supset \left(
\begin{array}{cccccc}
 1 & 0 & 0 & 0 & 0 & -1 \\
 0 & 1 & 0 & 0 & 0 & -1 \\
 0 & 0 & 1 & 0 & 0 & -1 \\
 0 & 0 & 0 & 1 & 0 & -1 \\
 0 & 0 & 0 & 0 & 1 & -1 \\
\end{array}
\right),
\end{equation}
and $\Delta$ has vertices
\begin{equation}
\Delta\supset\left(
\begin{array}{cccccc}
 5 & -1 & -1 & -1 & -1 & -1 \\
 -1 & 5 & -1 & -1 & -1 & -1 \\
 -1 & -1 & 5 & -1 & -1 & -1 \\
 -1 & -1 & -1 & 5 & -1 & -1 \\
 -1 & -1 & -1 & -1 & 5 & -1 \\
\end{array}
\right).
\end{equation}
If one considers the mirror Sextet, then $\Delta^\circ$ serves as the Newton polytope of the mirror sextet. Note that there is no other point in $\Delta^\circ\cap N$ than the vertices and the origin. Furthermore, all of the facets in $\Delta$ have five interior points. 

As expected, we find that all of the prime toric divisors in the Sextet have the hodge vector $h^\bullet_{\text{Sextet}}(D,\mathcal{O}_D)=(1,0,0,5).$ We also find that $h^{2,1}(D)$ for every prime toric divisor $D$ is $255.$ 

Unlike the Sextet, all of the prime toric divisors in the mirror Sextet are rigid and trivial in $h^{2,1}.$ Clearly, if one finds flux vacua with exponentially small vev of the GVW superpotential in the mirror Sextet, one can find $AdS_3$ vacua in M-theory.
\subsection{A fourfold uplift of $\Bbb{P}_{[1,1,1,6,9]}[18].$}
Now let us study an F-theory uplift $Y_1$ of the $\Bbb{P}_{[1,1,1,6,9]}[18]$ model.\footnote{For prior works on F-theory uplifts, see \cite{Sen:1996vd,Collinucci:2008zs,Collinucci:2009uh,Blumenhagen:2009up,Weigand:2018rez}.} We assign the weights to coordinates $u_i$'s as follows. $u_1,~u_2,~u_3$ have weight 1, $u_4$ has weight 6, and $u_5$ has weight 9. As a threefold base, we consider $B_3=\Bbb{P}_{[1,1,1,6]}\subset \Bbb{P}_{[1,1,1,6,9]},$ which is given by $\{u_5=0\}.$ Note that there is no reflexive polytope construction for $\Bbb{P}_{[1,1,1,6]},$ because 6 does not divide 9. This choice of the threefold base will lead to an F-theory uplift for the orientifold generated by $u_5\mapsto -u_5.$ Because $c_1(B)=c_1(\Bbb{P}_{[1,1,1,6,9]})/2,$ $\tilde{y}^2=b_2/576$ is formally equivalent to a weighted degree 18 polynomial which is symmetric under $\tilde{y}\mapsto -\tilde{y}.$

The $\Bbb{E}_8$ Weierstrass model can be understood as a single blow up of an anti-canonical class of $\Bbb{P}_{[1,1,1,6,18,27]}.$ The GLSM charge matrix for the single blow up is given by
\begin{center}
\begin{tabular}{ccccccc}
$u_1$ &$u_2$ &$u_3$&$u_4$ &$ x$&$y$&$ z$\\\hline
1&1&1&6&18&27&0\\\hline
0&0&0&0&2&3&1\\\hline
\end{tabular}
\end{center}
The GLSM for the $\Bbb{E}_7$ model is similarly given by
\begin{center}
\begin{tabular}{ccccccc}
$u_1$ &$u_2$ &$u_3$&$u_4$ &$ s$&$\tilde{y}$&$ z$\\\hline
1&1&1&6&18&9&0\\\hline
0&0&0&0&2&1&1\\\hline
\end{tabular}
\end{center}

The $\Bbb{E}_7$ model is described by a pair of the dual reflexive polytopes $\Delta^\circ\subset N$ and $\Delta\subset M$
\begin{equation}
\Delta^\circ\cap \Bbb{Z}^5\supset\left(
\begin{array}{cccccc}
 1 & 0 & 0 & 0 & 0 & -1 \\
 0 & 1 & 0 & 0 & 0 & -1 \\
 0 & 0 & 1 & 0 & 0 & -6 \\
 0 & 0 & 0 & 1 & 0 & -9 \\
 0 & 0 & 0 & 0 & 1 & -18 \\
\end{array}
\right),
\end{equation}
\begin{equation}
\Delta\cap\Bbb{Z}^5\supset
\left(
\begin{array}{cccccc}
 35 & -1 & -1 & -1 & -1 & -1 \\
 -1 & 35 & -1 & -1 & -1 & -1 \\
 -1 & -1 & -1 & -1 & 5 & -1 \\
 -1 & -1 & 3 & -1 & -1 & -1 \\
 -1 & -1 & -1 & 1 & -1 & -1 \\
\end{array}
\right).
\end{equation}

We record the combinatorial properties of the $\Delta^\circ$ in \S\ref{app:1}.\footnote{We have computed the combinatorial data using the {\tt{CYTools}} \cite{CYTools}.} We find that the 4d polytope $\Delta_{4d}$ generated as a convex hull of the vertices $\{v_1,v_2,v_3,v_4,v_6\}$ is the 4d reflexive polytope that defines $\Bbb{P}_{[1,1,1,6,9]}[18].$ There is a single point $v_7$ interior to a 2-face $\Theta_0^{(2)},$ which is a convex hull of points $v_1,~v_2,$ and $v_6.$ 

In this example, the genus of every vertex is non-trivial, meaning that the vertex prime toric divisors are not rigid. We compute the hodge vectors of the vertical vertex prime toric divisors
\begin{equation}
h^\bullet(D_{v_{1,2,6}},\mathcal{O}_{D_{v_{1,2,6}}})=(1,0,0,2),
\end{equation}
\begin{equation}
h^\bullet(D_{v_3},\mathcal{O}_{D_{v_3}})=(1,0,0,28),
\end{equation}
\begin{equation}
h^\bullet(D_{v_4},\mathcal{O}_{D_{v_4}})=(1,0,0,65).
\end{equation}
This result accords extremely well with the type IIB picture. Let us take as an example $D_1:=\{u_1=0\}$ in $\Bbb{P}_{[1,1,1,6,9]}[18].$ Because there are two additional toric coordinates $u_2$ and $u_3$ that are linearly equivalent to $u_1,$ we obtain $h^\bullet (D_1,\mathcal{O}_{D_1})=(1,0,2).$ Furthermore, the coordinates $u_2$ and $u_3$ are orientifold even. As a result, we obtain $h_-^{2}(D_1)=h_+^0(D_1,N_{D_1})=2$ which agrees exactly with the hodge vector $h^\bullet (D_{v_1})=(1,0,0,2).$

We now compute $h^{2,1}$ of the vertex divisors. As there is no 2 face $\Theta^{(2)}$ with $g(\Theta)l^*(\Theta)\neq0,$ $h^{2,1}$ of the vertex divisors do not depend on triangulations.
\begin{equation}
h^{2,1}(D_{v_{1,2,6}})=253,
\end{equation}
\begin{equation}
h^{2,1}(D_{v_3})=1758,
\end{equation}
\begin{equation}
h^{2,1}(D_{v_4})=2912.
\end{equation}

Similarly, we compute $h^\bullet (D_{v_7},\mathcal{O}_{D_{v_7}})=(1,0,0,0)$ and $h^{2,1}(D_{v_7})=0.$ Note that this divisor decends to a rigid prime toric divisor that hosts an O7-plane hence the non-higgsable cluster \cite{ftheory}. 

\subsection{GP orbifold of the uplift of $\Bbb{P}_{[1,1,1,6,9]}[18].$}
We now attempt to construct an orbifold $Y_2\equiv Y_1/G$ of $Y_1.$ The strategy is to extend the Green-Plesser group $G$ and its action to the Weierstrass model. Let us recall the Green-Plesser group $G$ and its action on $\Bbb{P}[1,1,1,6,9]$ \cite{Candelas:1994hw}
\begin{equation}
G\simeq \Bbb{Z}_6\times \Bbb{Z}_{18}:(u_1,u_2,u_3,u_4,u_5)\mapsto (\omega_{18} u_1,\omega_{18}^{-1}\omega_6 u_2,\omega_6^3 u_3,\omega_6^2u_4,u_5),
\end{equation}
where $\omega_i$ is an i-th root of unity. Conveniently, the Green-Plesser group does not act on $u_5.$ 

As in the previous sections, let $Y_2$ be the Calabi-Yau fourfold defined by the Netwon polytope $\Delta.$ In order to construct the F-theory compactification on $Y_2,$ we collect $G$ invariant monomials in $|Y_1|.$ Then, from the G-invariant subsets of $|Y_1|,$ we could learn a good deal about the Newton polytope for $Y_2.$ 

Let us summarize the findings regarding the G-invariant subset of $|Y_1|.$ There are in total 57 G-invariant monomials in $|Y_1|,$ and there are in total 13 non-trivial root automorphisms. Hence, we expect $\Delta_{Y_2}\cap \Bbb{Z}^5$ should contain 57 points, where one of them is the origin. Furthermore, among 57 points in $\Delta_{Y_2}\cap\Bbb{Z}^5,$ 13 points should lie inside the facets.

Given the fact that $\Delta_{11169}^\circ$ is contained in a hypersurface in the $N$ lattice, we propose an ansatz for $\Delta_{Y_2}^\circ$ as follow. First, let us consider the group action of $G$ on the lattice polytope $\Delta_{11169}^\circ$ and its containing lattice $N_{11169}.$ More explicitly, let us consider an isomorphism $N_{11169}\rightarrow \Bbb{Z}^4,$ such that $\Delta_{11169}^\circ\cap\Bbb{Z}^4$ has vertices
\begin{equation}
\Delta_{11169}^\circ\cap\Bbb{Z}^4\supset
\left(
\begin{array}{ccccc}
1&0&0&0&-1\\
0&1&0&0&-1\\
0&0&1&0&-6\\
0&0&0&1&-9
\end{array}
\right).
\end{equation}
Then, consider the group action of $G$ on $N_{11169}$ 
\begin{equation}
G\cdot v=\left(
\begin{array}{cccc}
 -1 & 17 & -1 & -1 \\
 -1 & -1 & -1 & -1 \\
 -1 & -1 & 2 & -1 \\
 -1 & -1 & -1 & 1 \\
\end{array}
\right)^T\cdot v,
\end{equation}
which maps the vertices of $\Delta_{11169}^\circ$ to
\begin{equation}
g\cdot (\Delta_{11169}\cap\Bbb{Z}^4) \supset
\left(
\begin{array}{ccccc}
 -1 & -1 & -1 & -1 & 17 \\
 17 & -1 & -1 & -1 & -1 \\
 -1 & -1 & 2 & -1 & -1 \\
 -1 & 1 & -1 & -1 & -1 \\
\end{array}
\right).
\end{equation}
Now consider an embedding of $N_{11169}$ into $N$ and an extension of the group action on $u\in N$ such that
\begin{equation}
g\cdot u=\left(
\begin{array}{ccccc}
 -1 & 17 & -1 & -1 &0\\
 -1 & -1 & -1 & -1 &0\\
 -1 & -1 & 2 & -1 &0\\
 -1 & -1 & -1 & 1 &0\\
 0&0&0&0&1\\
\end{array}
\right)^T\cdot u.
\end{equation}
Given the group action of $G$ on $N,$ we now obtain a candidate lattice polytope $\Delta_G^\circ\cap \Bbb{Z}^5=G\cdot (\Delta^\circ\cap \Bbb{Z}^5),$ which has vertices
\begin{equation}
\Delta_{G}^\circ\cap\Bbb{Z}^5\supset
\left(
\begin{array}{cccccc}
 17 & 0 & -1 & -1 & -1 & -1 \\
 -1 & 0 & 17 & -1 & -1 & -1 \\
 -1 & 0 & -1 & 2 & -1 & -1 \\
 -1 & 0 & -1 & -1 & 1 & -1 \\
 -18 & 1 & 0 & 0 & 0 & 0 \\
\end{array}
\right).
\end{equation}
The dual polytope $\Delta_G,$ which is meant to be the Newton polytope, contains vertices
\begin{equation}
\Delta_{G}\cap\Bbb{Z}^5\supset
\left(
\begin{array}{cccccc}
 -1 & 1 & -1 & 1 & -1 & -3 \\
 0 & 0 & 0 & 0 & 2 & -2 \\
 0 & 0 & 2 & 0 & 0 & -12 \\
 2 & 0 & 0 & 0 & 0 & -18 \\
 -1 & 1 & -1 & -1 & -1 & -1 \\
\end{array}
\right).
\end{equation}

Let us check if our ansatz for $\Delta_G$ and $\Delta_G^\circ$ is consistent with what we have found about the G-invariant subset of $|Y_1|.$ What we find is that the monomials constructed from $\Delta_G^\circ$ and $\Delta_G$ are one to one to the monomials in the G-invariant subset of $|Y_1|,$ and the root automorphisms also map from one to the other via a one to one map. Hence, we conclude $\Delta_G^\circ$ is isomorphic to $\Delta_{Y_2}^\circ$ and likewise $\Delta_G$ is isomorphic to $\Delta_{Y_2}.$ We record the combinatorial data in \S\ref{app:2}.

Let us study the fourfold $Y_2$ in more detail. Among 475 prime toric divisors, only three prime toric divisors have non-trivial hodge vectors.\footnote{This can be easily seen from the fact that $g(\Theta)l^*(\Theta)=0$ for all faces except for $v_2,~v_4,~v_5.$} Here is a list of three non-trivial hodge vectors
\begin{equation}
h^\bullet(D_{v_2},\mathcal{O}_{D_{v_2}})=(1,0,0,10),
\end{equation}
\begin{equation}
h^\bullet(D_{v_4},\mathcal{O}_{D_{v_4}})=(1,0,0,1),
\end{equation}
\begin{equation}
h^\bullet(D_{v_5},\mathcal{O}_{D_{v_5}})=(1,0,0,2).
\end{equation}

We also find that all of the rigid prime toric divisors are trivial in $h^{2,1}.$ It can be easily checked that no rigid divisor is in a convex hull of $v_2,~v_4,$ and $v_5,$ which is the unique 2-face with non-trivial genus. Hence, all of the rigid edge-divisors and 2-face-divisors are trivial in $h^{2,1}.$ Let us therefore compute $h^{2,1}$ of $D_{v_{1,2,6}}.$ Because there is a $\Bbb{Z}_3$ symmetry that relates $v_{1,2,6},$ without loss of generality, we will study $v_1$ explicitly and draw conclusions for $v_2$ and $v_6$ as well. Recall that $h^{2,1}(D_{v_1}),$ c.f. \eqref{res1}, is written as 
\begin{align}
h^{2,1}(D_{v_1})=&\varphi_2(v_1^\circ)-\sum_{e\geq v_1} l^*(f^\circ)\\
=&l^*(2v_1^\circ)-5 l^*(v_1^\circ)-\sum_{e\geq v_1} l^*(f^\circ).
\end{align}
Let us compute $l^*(2v_1^\circ).$ $2v_1^\circ$ is a convex hull of the following points
\begin{equation}
2v_1^\circ\supset
\left(
\begin{array}{ccccc}
 2 & -2 & -2 & -2 & -6 \\
 0 & 4 & 0 & 0 & -4 \\
 0 & 0 & 4 & 0 & -24 \\
 0 & 0 & 0 & 4 & -36 \\
 2 & -2 & -2 & -2 & -2 \\
\end{array}
\right).
\end{equation}
It is straightforward to verify that there is no interior point in $2v_1^\circ,$ hence we obtain $l^*(2v_1^\circ)=0.$ Because $h^{2,1}$ cannot be negative, $l^*(v_1^\circ)$ and $\sum_{f\geq v_1}l^*(f^\circ)$ should vanish. Let us verify this claim explicitly. As was computed in \S\ref{app:2}, $D_{v_1}$ has no deformation modulus and $l^*(v_1^\circ)=0.$ Furthermore, all the 2-faces containing $v_1$ have trivial genus so that $\sum_{f\geq v_1}l^*(f^\circ)=0.$  As a result, we obtain $h^{2,1}(D_{v_{1,2,6}})=0.$

A comment is in order. Albeit this F-theory compactification is attractive, it is not clear how to extract the j-invariant of $Y_2$ or find the weakly coupled type IIB description. After an MPCP desingularization, the section $X^4$ in the defining equation of the $\Bbb{E}_7$ model, $Y_2,$ becomes $X^4\prod_i \epsilon_i^2,$ where $\epsilon_i$ are some homogeneous coordinates. Note that this is not the unique feature of the $\Bbb{E}_7$ model, in fact the Weierstrass form suffers from the similar problem. Hence, we no longer have the canonical form of the elliptic fibration which complicates the analysis. It will be interesting to compute the j-invariant of this model in the future.

\section*{Acknowledgements}
The author thanks Andres Rios-Tascon, Andreas P. Braun, Jakob Moritz, and Liam McAllister for useful discussions and comments on the draft. The work of the author was supported in part by NSF grant PHY-1719877.
\newpage
\begin{appendix}
\addtocontents{toc}{\protect\setcounter{tocdepth}{1}}
\section{Review}\label{S:review}
In this section, we review the relevant results of \cite{reid1983decomposition,danilov1987newton,Batyrev:1994hm,Klemm:1996ts,cox2011toric,Braun:2016igl,Braun:2017nhi}. 

\subsection{Toric varieties and stratification}
Let $M$ be an abelian group of rank n, one can think of $M$ as an integral lattice of dimension $n$ under the isomorphism $i:M\rightarrow \Bbb{Z}^n.$ We consider a polytope $\Delta,$ which is integral with respect to $M.$ We further require that $\Delta$ is reflexive. We will regard this reflexive polytope $\Delta$ as the Newton polytope for the Calabi-Yau manifold in the latter. One can construct a toric variety through the Proj construction of the graded rings generated by cones over faces of $\Delta.$ In this note, we do not consider the Proj construction but rather the normal fan construction which will be described in the subsequent paragraphs.

Consider a dual of $M,$ $N=\text{Hom}(M,\Bbb{Z}).$ Likewise, we take a polar dual $\Delta^\circ$ of the lattice polytope $\Delta.$ Given the dual lattice and the polytope data, we construct the toric fan as follows. Let $\Theta^{(l)}$ be an $l$-dimensional face of $\Delta.$ Then, we define a convex $n$-dimensional cone $\sigma^\vee(\Theta^{(l)}) \subset M\otimes \Bbb{Q}$ which is a set of vectors $\lambda (p-p')$ where $\lambda\in\Bbb{Q}_{\ge0},$ $p\in\Delta,$ $p'\in \Theta^{(l)}.$ We define $\sigma(\Theta^{(l)})$ te be the $(n-l)$ dimensional dual cone of $\sigma^\vee (\Theta^{(l)}).$ Let $N(\sigma(\Theta^{(l)})$ be a minimal $n-l$ dimensional sublattice of $N$ containing $\sigma\cap N.$ Then, for each cone $\sigma$ of dimension d, we define an $n$ dimensional \emph{affine} toric varitey $\Bbb{A}_{\sigma,N}=\text{Spec}[\sigma^\vee \cap M]$ which is isomorphic to $(C^*)^{n-d}\times \Bbb{A}_{\sigma,N(\sigma)}.$ The toric fan $\Sigma$ is the polyhedral fan of cones $\sigma.$ We define $\Sigma^{(i)}$ be the set of all $i$-dimensional cones in $\Sigma,$ and $\Sigma^{[i]}$ be the subfan of $\Sigma$ that contains all cones of dimension less than or equal to $i.$ By gluing the affine toric varities $\Bbb{A}_{\sigma,N}$ for all $\sigma$ canonically, we obtain the n-dimensional toric variety $\Bbb{P}_{\Sigma,N}.$

We now present a different, and quite useful, angle on the construction of $\Bbb{P}_{\Sigma,N},$ the stratification. An $n$-dimensional toric variety can be understood as a compactification of an $n$-dimensional torus $(C^*)^n.$ The first step of the compactification is proceeded by attaching normal crossing divisors, each of which irreducible components is $(C^*)^{n-1},$ to $(C^*)^n.$ At this stage, the compactification is not completed because the normal crossing divisors used to compactify $(C^*)^n$ themselves are not compact. Hence, one then sequentially compactifies $(C^*)^{n-d}$ by attaching $(C^*)^{n-d-1}$ until $d=n-1.$ We call $(C^*)^d$ a d-dimensional strata of $\Bbb{P}_{\Sigma}.$ An elementary example is the toric variety $\Bbb{P}^1,$ which can be constructed by attaching two points to a one dimensional algebraic torus $C^*.$

The stratification procedure may look quite distant from the toric fan construction. But, in fact, the toric fan construction provides a natural way to stratify toric varieties. We recall the following proposition \cite{Batyrev:1994hm}
\begin{proposition}\label{prop1}
Let $\Delta$ be an $n$-dimensional $M$-integral polytope in $M\otimes \Bbb{Q},$ $\Sigma=\Sigma(\Delta)$ the corresponding complete rational polyhedral fan in $N\otimes \Bbb{Q}.$ Then
\begin{itemize}
\item For any face $\Theta\subset \Delta,$ the affine toric variety $\Bbb{A}_{\sigma(\Theta),N}$ is the minimal $\Bbb{T}$-invariant affine open subset in $\Bbb{P}_{\Sigma}$ containing the $\Bbb{T}$-orbit $\Bbb{T}_{\Theta}.$
\item Let $\Bbb{T}_\sigma:=\Bbb{T}_{\sigma(\Theta)}.$ There exists a one-to-one correrspondence between $s$-dimensional cones $\sigma\in\Sigma$ and $(n-s)$-dimensional $\Bbb{T}$-orbits $\Bbb{T}_\sigma$ such that $\Bbb{T}_{\sigma'}$ is contained in the closure of $\Bbb{T}_\sigma$ iff $\sigma$ is a face of $\sigma'.$
\item $\Bbb{P}_{\Sigma}^{[i]}=\cup_{\dim \sigma\leq i}\Bbb{T}_{\sigma}$ is an open $\Bbb{T}$-invariant subvariety in $\Bbb{P}_{\Sigma}=\Bbb{P}_\Delta,$ and $\Bbb{P}_{\Sigma}\backslash \Bbb{P}_{\Sigma}^{[i]}=\Bbb{P}_\Delta^{(i)}.$ 
\end{itemize}
\end{proposition}
This proposition also provides a very intuitive understanding on the nature of the points in the polytope $\Delta^\circ.$ To each vertex $v_i\in \Delta^\circ,$ we associate a homogeneous coordinate $x_i,$ and $\Bbb{P}_{v^\circ}.$ $\Bbb{P}_{v^\circ}$ is a compacitification of $(C^*)^{n-1}.$ Note that the toric variety $\Bbb{P}_\Sigma$ can have various singularities, we comment on the singularity in the next section. 

\subsection{Comments on singularities}

The singularities of the toric variety $\Bbb{P}_{\Sigma}$ are determined by the integrality of the faces in $\Delta^\circ,$ or equivalently the integrality of the toric fan \cite{reid1983decomposition,Batyrev:1994hm}. In general, there is a face $\Theta\in\Delta^\circ$ that is not integral with respect to the lattice $N$ which then indicates a singularity. This kind of singularities can be resolved by blowing up along the singular subvarieties. In the toric fan construction, such blow-ups can be carried out by a refinement of the toric fans which naturally gives a star triangulation of the reflexive polytope $\Delta^\circ.$ In this section, we review which types of triangulations are good enough so that we can compute the hodge numbers reliably. We also comment on in which cases the singularities are completely resolved.

First, let us define $p_\sigma$ to be the unique $\Bbb{T}/\Bbb{T}_\sigma$-invariant point on the $d$-dimensional affine toric variety $\Bbb{A}_{\sigma,N(\sigma)}.$ It suffices to study the singularity around $p_\sigma$ for $\sigma\in \Sigma$ as the algebraic torus is locally indistinguishable. We then recall a proposition \cite{reid1983decomposition}
\begin{proposition}
Let $n_1,\dots,n_r\in N$ ($r\geq s$) be primitive $N$-integral generators of all 1-dimensional faces of an $d$-dimensional cone $\sigma.$
\begin{itemize}
	\item The point $p_\sigma\in \Bbb{A}_{\sigma,N(\sigma)}$ is $\Bbb{Q}$-factorial (or quasi-smooth) if and only if the cone $\sigma$ is simplicial.
	\item The point $p_\sigma\in \Bbb{A}_{\sigma,N(\sigma)}$ is $\Bbb{Q}$-Gorenstein if and only if the elements $n_1,\dots,n_r$ are contained in an affine hyperplane
	\begin{equation}
	H_\sigma : \{ y\in N_{\Bbb{Q}}|\langle k_\sigma,y\rangle =1\}\,,
	\end{equation}
	for some $k_\sigma\in M_{\Bbb{Q}}.$ Moreover, $\Bbb{A}_{\sigma,N(\sigma)}$ is Gorenstein if and only if $k_\sigma\in M.$
\end{itemize}
\end{proposition}

We first discuss which faces constitute good faces. A d-dimensional simplex $\sigma\in \Bbb{Q}^n$ with vertices in $\Bbb{Z}^n$ is called $P$ elementary if the minimal $d$-dimensional affine $\Bbb{Q}$ lattice $A(\sigma)$ satisfies following property: $\sigma\cap(A(\sigma)\cap \Bbb{Z}^n)$ contains only the vertices of $\sigma.$ $\sigma$ is called regular, if $\sigma$ is unimodular with respect to $A(\sigma)\cap \Bbb{Z}^n.$ Note in particular that every elementary simplex of dimension less than or equal to 2 is regular. We then recall a very important proposition \cite{Batyrev:1994hm}
\begin{proposition}
Let $\Bbb{P}_\Sigma$ be a toric variety with only $\Bbb{Q}$-Gorenstein singularities. Then
\begin{itemize}
	\item $\Bbb{P}_\Sigma$ has only $\Bbb{Q}$-factorial terminal singularities if and only if for every cone $\sigma\in\Sigma$ the polyhedron
	\begin{equation}
	P_\sigma=\sigma \cap \{ y\in\Bbb{N}_{\Bbb{Q}}|\langle k_\sigma,y\rangle\leq 1\}
	\end{equation}
	is elementary.
	\item $\Bbb{P}_\Sigma$ is smooth if and only if for every cone $\sigma\in\Sigma$ the polyhedron $\Bbb{P}_\sigma$ is regular.
\end{itemize}
Note that in this note we will only consider $\Bbb{P}_\Sigma$ with only Gorenstein singularities.
\end{proposition}
As a corollary, if $\Bbb{P}_\Sigma$ has only Gorenstein $\Bbb{Q}$-factorial terminal singularities, then the open toric subvarieties $\Bbb{P}_\Sigma^{[3]}$ is smooth. This corollary is a central result that guarantees the smoothness of the subvarieties up to co-dimension three under a Maximal Projective Crepant Partial desingularization (MPCP-desingularization), which is a crepant resolution of the singularities that only leave out $Q$-factorial terminal singularities unresolved. Furthermore, in \cite{Batyrev:1994hm} for any $\Bbb{P}_\Sigma$ with only Gorenstein singularities it was shown that there exists at least one MPCP-desingularization, and each MPCP desingularization is defined by a fine regular star triangulation of $\Delta^\circ.$\footnote{See, for example, \cite{Demirtas:2018akl,Demirtas:2020dbm} for detailed discussions on FRSTs.}

A few comments are in order. In general, there is no reason to expect that the simplices induced by an MPCP will all be regular. Hence, for a general MPCP desingularization, the toric variety $\Bbb{P}_\Sigma$ will still suffer from $\Bbb{Q}$-factorial terminal singularities. This result implies that any toric hypersurface CY 3-folds can be made completely smooth via MPCPs, while toric hypersurface CY 4-folds may still have some singularities. It is nevertheless very important not to think that there is no complete resolution of the singularities in toric hypersurface CY 4-folds. If there is an FRST $\mathcal{T}$ of $\Delta^\circ$ such that a proper subset of the simplices are regular, the CY 4-folds in $\Bbb{P}_{\Delta^\circ}$ under $\mathcal{T}$ will be smooth.

%

\subsection{Calabi-Yau hypersurfaces}
A $d$-dimensional Calabi-Yau manifold can be embedded into a $d+1$-dimensional toric variety $\Bbb{P}_{\Sigma},$ on which an MPCP desingularization was carried out, as a vanishing locus of a section of the anti-canonical line bundle $-K_{\Bbb{P}_\Sigma}.$ Because the first Chern class of the anti-canonical line bundle is a formal sum of the divisors, we obtain
\begin{equation}
c_1(-K_{\Bbb{P}_\Sigma})=\sum_{\nu_i \in \Sigma(1)}D_i.
\end{equation}
The corresponding Newton polytope for the anti-canonical line bundle is then
\begin{equation}
\Delta_{CY}:=\{ m\in M|\langle m,\nu\rangle \geq -1\forall\nu\in \Sigma(1)\},
\end{equation}
which is the polar dual of $\Delta^\circ.$ Hence, we conclude that $\Delta=\Delta_{CY}$ and a point $p\in\Delta$ defines a section
\begin{equation}
\mathcal{F}_p=a_p \prod_{\nu_i\in\Sigma(1)} z_i^{\langle p,\nu_i\rangle +1}\,.
\end{equation}
Note that points $\nu\in\Sigma(1)$ coincide with points in $(\mathcal{T}\cap \Delta^\circ)\cap \Bbb{N}.$ By $\mathcal{F},$ we denote a generic section of $-K_{\Bbb{P}_\Sigma}$
\begin{equation}
\mathcal{F}=\sum_{p\in \Delta}a_p \prod_{\nu_i\in\Sigma(1)} z^{\langle p,\nu_i\rangle +1}_i \,.
\end{equation}

Let $\overline{Z}_{\mathcal{F},\Delta}$ be a vanishing locus of $\mathcal{F}$ in $\Bbb{P}_{\Sigma}.$ The bar above $Z_{\mathcal{F},\Delta}$ is to denote that $\overline{Z}_{\mathcal{F},\Delta}$ is a compactification of $Z_{\mathcal{F},\Delta}\in (C^*)^{d+1}.$ Assuming that the hypersurface $\overline{Z}_{\mathcal{F},\Delta}$ or in a shorthanded notation $\overline{Z}_\mathcal{F}$ is regular, meaning that for every face $\Theta\subset\Delta$ the affine variety $Z_{\mathcal{F},\Theta}=\overline{Z}_{\mathcal{F},\Delta}\cap \Bbb{T}_{\Theta}$ is empty or a smooth hypersurface in $\Bbb{T}_\Theta,$ we obtain a stratification
\begin{equation}
\overline{Z}_{\mathcal{F},\Delta}=\coprod_{\Theta\subset\Delta}Z_{\mathcal{F},\Theta}\,.
\end{equation}
Equivalently, we define $\overline{Z}_{\mathcal{F},\Sigma}:=\overline{Z}_{\mathcal{F},\Delta}$ and $ Z_{\mathcal{F},\sigma(\Theta)}:=Z_{\mathcal{F},\Theta}.$ We define an open subvariety $Z_{\mathcal{F},\Sigma}^{[i]}\subset \overline{Z}_{\mathcal{F},\Sigma}$ as
\begin{equation}
Z_{\mathcal{F},\Sigma}^{[i]}:=\overline{Z}_{\mathcal{F},\Sigma}\cap \Bbb{P}_{\Sigma}^{[i]}\,.
\end{equation}
$\mathcal{F}$
In the beginning of this section, we assumed that an MPCP desingularization for the toric fan was carried out. Now we discuss the implication of an MPCP desingularization. An MPCP desingularization is defined by a refinement $\phi:\Sigma\rightarrow \Sigma(\Delta)$ of the toric fan $\Sigma(\Delta).$ The refinement procedure introduces new exceptional divisors whose topologies are dictated by
\begin{equation}
Z_{ \mathcal{F},\sigma} \equiv Z_{ \phi_*(\mathcal{F}),\sigma'}\times(C^*)^{\dim \sigma'-\dim\sigma}\,,
\end{equation}
for $\sigma\in\Sigma$ and $\phi(\sigma)\subset \sigma'\in \Sigma(\Delta).$ This result will be used in an important manner when we prove the combinatorial formulas for $h^{2,1}$ of divisors.

Borrowing the results on the singularities of $\Bbb{P}_{\Sigma}^{[i]},$ we now recall the singularities of the stratas of $\overline{\Bbb{Z}}_{\mathcal{F}}$ \cite{Batyrev:1994hm}.
\begin{proposition}
For any regular hypersurface $\overline{Z}_{\mathcal{F},\Sigma}\subset\Bbb{P}_{\Sigma},$ the open subset $Z_{\mathcal{F},\Sigma}^{[1]}$ consists of smooth points of $\overline{Z}_{\mathcal{F},\Sigma}.$ Moreover,
\begin{itemize}
\item $Z_{\mathcal{F},\Sigma}^{[2]}$ consists of smooth points if $\Bbb{P}_{\Sigma}$ has only terminal singularities.
\item $Z_{\mathcal{F},\Sigma}^{[3]}$ consists of smooth points if $\Bbb{P}_{\Sigma}$ has only $\Bbb{Q}$-factorial Gorenstein terminal singularities.
\item $Z_{\mathcal{F},\Sigma}^{[d]}=\overline{Z}_\mathcal{F}$ is smooth iff $\Bbb{P}_\Sigma^{[d]}$ is smooth.
\end{itemize}
\end{proposition}
As a result, given an arbitrary FRST $\mathcal{T},$ at the worst there are singularities in $\Bbb{Z}_{\mathcal{F},\Sigma}^{[4]}\backslash\Bbb{Z}_{\mathcal{F},\Sigma}^{[3]}$ for a CY 4-fold embedded in a toric fivefold. As it is well known, a CY 3-fold is completely smooth as an embedding to a toric fourfold.

\subsection{Hodge-Deligne numbers}
In the previous sections, we introduced stratifications for an arbitrary toric Calabi-Yau manifold. The stratification for the Calabi-Yau manifold will come in handy when computing the hodge structure of the prime toric divisors, as the stratifications of the prime toric divisors descend from the stratification of the Calabi-Yau manifold. For more inclined readers, we recommend \cite{voisin2002hodge,voisin2002hodge2}.

In order to compute the hodge structure algorithmically, we wish to have a character $e$ that satisfies good properties, a few of which are as follows. First, for a disjoint union $X=\coprod X_i,$ we want the character to satisfy $e(X)=\sum e(X_i).$ Second, for a compact quasi-smooth variety, we want the character $e^{p,q}$ to hold equivalent information as $h^{p,q}.$ Phrased differently, we require that $e^{p,q}$ computes the pure hodge structure. Third, we want the Kunneth isomorphism compatible with the hodge structure to exist for the character $e.$ Once those conditions are met, we can utilize the character $e$ to determine the hodge structures of the prime toric divisors from the stratification data quite algorithmically.

The topological invariant of interest is the Hodge-Deligne number which is defined as
\begin{equation}
e^{p,q}(X)=\sum_k (-1)^k h^{p,q}\left( H_c^k(X)\right)\,,
\end{equation}
where $H_c^k(X)$ is the dimension $k$ cohomology with compact support. We then define the Hodge-Delign character as
\begin{equation}
e(X)=\sum_{p,q}e^{p,q}(X)x^p \bar{x}^q\,.
\end{equation}
The Hodge-Deligne character and number satisfy various useful properties, and we highlight some of them.
\begin{itemize}
\item For a compact quasi-smooth variety $X,$ $e^{p,q}(X)=(-1)^{p+q}h^{p,q}(X).$
\item $h^{p,q}(H_c^k(X))=0$ for $p+q>k.$
\item Let $X$ be a disjoint union of $X_i$'s such that $X=\coprod X_i.$ Then $e(X)=\sum_i e(X_i).$
\item For a fiber bundle $f:X\rightarrow Y$ with $f(pt)^{-1}=F$ that is locally trivial in the Zariski topology, $e(X)=e(Y)\times e(F).$
\item Let $X$ be a smooth variety, and $\overline{X}$ be a compactification such that $D:=\overline{X}\backslash X$ is a normal crossing divisor in $\overline{X}.$ Then, $e^p(X):=\sum_q e^{p,q}(X)=(-1)^{p}\chi(\overline{X},\Omega_{(\overline{X},D)}^{p}).$
\end{itemize}

To compute the Hodge-Deligne numbers of toric hypersurfaces of various dimensions, we recall three important ingredients.\footnote{Note that we will not restrict ourselves to the reflexive polytopes, because the prime toric divisor's strata contains such hypersurfaces that are not of Calabi-Yaus.}
\begin{itemize}
\item There is a Gysin homomorphism $H^i(\overline{Z}_\Theta,\Bbb{C})\rightarrow H^{i+2}(\Bbb{P}_\Theta,\Bbb{C})$ which is an isomorphism for $i>\dim \overline{Z}_\Theta$ and is a surjection for $i=\dim\overline{Z}_\Theta.$
\item The following short sequence is exact
\begin{equation}
0\rightarrow \Omega^p_{(\overline{Z}_\Theta,D_Z)}\otimes_{\mathcal{O}_{\Bbb{P}_\Theta}}\mathcal{O}_{\Bbb{P}_{\Theta}}(-\overline{Z}_{\Theta})\rightarrow \Omega^{p+1}_{(\Bbb{P}_{\Theta},D)}\otimes_{\mathcal{O}_{\Bbb{P}_\Theta}}\mathcal{O}_{\overline{Z}_\Theta}\rightarrow \Omega^{p+1}_{(\overline{Z}_{\Theta},D_Z)}\rightarrow 0\,.\label{exact1}
\end{equation}
\item For the toric variety $\Bbb{P}_{\Theta}$ and its hypersurface $\overline{Z}_\Theta,$ the fundamental exact sequence exists
\begin{equation}
0\rightarrow \mathcal{O}_{\Bbb{P}_{\Theta}}(-\Theta)\rightarrow \mathcal{O}_{\Bbb{P}_\Theta}\rightarrow \mathcal{O}_{\overline{Z}_\Theta}\rightarrow 0\,.\label{exact2}
\end{equation}
\end{itemize}

Let $n$ be $\dim \Bbb{P}_\Theta.$ From the existence of the Gysin isomorphism, one can compute for $p+q>n-1$
\begin{equation}
e^{p,q}(Z_\Theta)=e^{p+1,q+1}(\Bbb{T}_\Theta)\,,
\end{equation}
which is equal to $(-1)^{n+p+1}\binom{n}{p+1}$ for $p=q$ and is equal to $0$ otherwise. Furthermore, from the exact sequences \eqref{exact1} and \eqref{exact2}, we obtain
\begin{equation}
(-1)^{n-1}e^{p}(Z_\Theta)=(-1)^p \binom{n}{p+1}+\varphi_{n-p}(\Theta)\,,
\end{equation}
where $\varphi_i(\Theta):= (-1)^{i}\sum_{j\geq 1} (-1)^j \binom{n+1}{i-j} l^*(j\Theta).$ Finally, we relate the Hodge-Deligne numbers of $\overline{Z}_\Theta$ to the Hodge-Deligne numbers of $Z_\Gamma$ for $\Gamma\leq \Theta$
\begin{equation}
e^{p,q}(\overline{Z}_\Theta)=\sum_{\Gamma\leq \Theta}e^{p,q}(Z_\Gamma)\,.
\end{equation}
Because the Hodge-Deligne numbers $e^{p,q}(Z_\Theta)$ are known for $p+q>n-1$ and the Poincare duality requires $e^{p,q}(\overline{Z}_\Theta)=e^{n-p-1,n-q-1}(\overline{Z}_\Theta),$ we have determined all of the Hodge-Deligne numbers. In the rest of the section, we collect some of the Hodge-Deligne numbers for $k$-dimensional polytopes $\Theta^{(k)}$ up to dimension 4.

Let $l^i(\Theta)$ be the number of points in $i$-dimensional simplex in $\Theta$ and $l^*(\Theta)$ be the number of interior points in $\Theta.$ Then we obtain the following identities.
\begin{equation}
e^{0,0}(Z_{\Theta^{(k)}})=(-1)^{k-1}(l^1(\Theta^{(k)})-1)\,.
\end{equation}
\begin{equation}
e^{i,j}(\Theta^{(2)})=
\begin{tabular}{|c|c|}
\hline
$-l^*(\Theta^{(2)})$&1\\
\hline
$1-l^1(\Theta^{(2)})$&$-l^*(\Theta^{(2)})$\\
\hline
\end{tabular}\,.
\end{equation}
\begin{equation}
e^{i,j}(\Theta^{(3)})=\begin{tabular}{|c|c|c|}
\hline
$l^*(\Theta^{(3)})$&0&1\\
\hline
$l^2(\Theta^{(3)})-l^1(\Theta^{(3)})$&$ e^{1,1}(\Theta^{(3)})$&0\\
\hline
$l^1(\Theta^{(3)})-1$& $l^2(\Theta^{(3)})-l^1(\Theta^{(3)})$ & $l^*(\Theta^{(3)})$\\
\hline
\end{tabular}\,,
\end{equation}
where $e^{1,1}(\Theta^{(3)})=\varphi_2(\Theta^{(3)})-l^2(\Theta^{(3)})+l^1(\Theta^{(3)})-3. $ For $e^{i,j}(\Theta^{(4)})$ we obtain
\begin{equation}
\begin{tabular}{|c|c|c|c|}
\hline
$-l^*(\Theta^{(4)})$&0&0&1\\
\hline
$-l^3(\Theta^{(4)})+l^2(\Theta^{(4)})$&$ e^{2,1}(\Theta^{(4)})$&-4&0\\
\hline
$-l^2(\Theta^{(4)})+l^1(\Theta^{(4)})$&$ e^{1,1}(\Theta^{(4)})$&$e^{2,1}(\Theta^{(4)})$&0\\
\hline
$1-l^1(\Theta^{(4)}) $& $-l^2(\Theta^{(4)})+l^1(\Theta^{(4)})$&$-l^3(\Theta^{(4)})+l^2(\Theta^{(4)})$&$-l^*(\Theta^{(4)})$\\
\hline
\end{tabular}
\end{equation}
where $e^{2,1}(\Theta^{(4)})=l^3(\Theta^{(4)})-l^2(\Theta^{(4)})-\varphi_2(\Theta^{(4)})$ and $e^{1,1}=-\varphi_3(\Theta^{(4)})+\varphi_2(\Theta^{(4)})-l^3(\Theta^{(4)})+2l^2(\Theta^{(4)})-l^1(\Theta^{(4)})+6.$

\section{Combinatorial data for the Calabi-Yau fourfolds.}
\subsection{Combinatorial data for the uplift of $\Bbb{P}_{[1,1,1,6,9]}[18].$}\label{app:1}

\begin{center}
\begin{tabular}{|c||c|c|c|}
\hline
dim & face &  $\#$ int pts & genus \\
\hline
0 & 1 & 1 & 2\\
0 & 2 & 1 & 2\\
0 & 3&1&28\\
0&4&1&65\\
0&5&1&376\\
0&6&1&2 \\
\hline
1&1,2&0&1\\
1&1,3&0&8\\
1&1,4&0&16\\
1&1,5&0&62\\
1& 1,6& 0 &1\\
1&2,3&0&8\\
1&2,4&0&16\\
1&2,5&0&62\\
1&2,6&0&1\\
1&3,4&0&136\\
1&3,5&0&489\\
1&3,6 &0&8\\
1&4,5&0&860\\
1&4,6&0&16\\
1&5,6&0&62\\
\hline
2&1,2,3&0&1\\
2&1,2,4&0&2\\
2&1,2,5&0&7\\
2&1,2,6&1&0\\
2&1,3,4&0&17\\
2&1,3,5&0&51\\
2& 1,3,6 & 0&1\\
2&1,4,5&0&85\\
2&1,4,6&0&2\\
2&1,5,6&0&7\\
\hline
\end{tabular}
\qquad
\begin{tabular}{|c||c|c|c|}
\hline
dim & face &  $\#$ int pts & genus \\
\hline
2&2,3,4&0&17\\
2&2,3,5&0&51\\
2&2,3,6&0&1\\
2&2,4,5&0&85\\
2&2,4,6&0&2\\
2&2,5,6&0&7\\
2&3,4,5&0&595\\
2&3,4,6&0&17\\
2&3,5,6&0&51\\
2&4,5,6&0&85\\
\hline
3&1,2,3,4&0&1\\
3&1,2,3,5&0&3\\
3&1,2,3,6&2&1\\
3&1,2,4,5&0&5\\
3&1,2,4,6&1&1\\
3&1,2,5,6&0&1\\
3&1,3,4,5&0&35\\
3&1,3,4,6&0 &1\\
3&1,3,5,6&0&3\\
3&1,4,5,6&0&5\\
3&2,3,4,5&0&35\\
3&2,3,4,6&0&1\\
3&2,3,5,6&0&3\\
3&2,4,5,6&0&5\\
3&3,4,5,6&0&35\\
\hline
4&1,2,3,4,5&0&0\\
4&1,2,3,5,6&0&0\\
4&1,2,3,4,6&1&0\\
4&1,2,4,5,6&0&0\\
4&1,3,4,5,6&0&0\\
4&2,3,4,5,6&0&0\\
\hline

\end{tabular}
\end{center}

\subsection{Combinatorial data for $Y_2.$}\label{app:2}
\begin{center}
\begin{tabular}{|c||c|c|c|}
\hline
dim & face &  $\#$ int pts & genus \\
\hline
0 & 1 & 1 & 0\\
0 & 2 & 1 & 10\\
0 & 3&1&0\\
0&4&1&1\\
0&5&1&2\\
0&6&1&0 \\
\hline
1&1,2&0&0\\
1&1,3&17&0\\
1&1,4&2&0\\
1&1,5&1&0\\
1& 1,6& 17 &0\\
1&2,3&0&0\\
1&2,4&0&6\\
1&2,5&0&11\\
1&2,6&0&0\\
1&3,4&2&0\\
1&3,5&1&0\\
1&3,6 &17&0\\
1&4,5&0&1\\
1&4,6&2&0\\
1&5,6&1&0\\
\hline
2&1,2,3&0&0\\
2&1,2,4&0&0\\
2&1,2,5&0&0\\
2&1,2,6&0&0\\
2&1,3,4&16&0\\
2&1,3,5&8&0\\
2& 1,3,6 & 136&0\\
2&1,4,5&1&0\\
2&1,4,6&16&0\\
2&1,5,6&8&0\\
\hline
\end{tabular}
\qquad
\begin{tabular}{|c||c|c|c|}
\hline
dim & face &  $\#$ int pts & genus \\
\hline
2&2,3,4&0&0\\
2&2,3,5&0&0\\
2&2,3,6&0&0\\
2&2,4,5&0&4\\
2&2,4,6&0&0\\
2&2,5,6&0&0\\
2&3,4,5&1&0\\
2&3,4,6&16&0\\
2&3,5,6&8&0\\
2&4,5,6&1&0\\
\hline
3&1,2,3,4&0&1\\
3&1,2,3,5&0&1\\
3&1,2,3,6&0&1\\
3&1,2,4,5&0&1\\
3&1,2,4,6&0&1\\
3&1,2,5,6&0&1\\
3&1,3,4,5&2&1\\
3&1,3,4,6& 65 &1\\
3&1,3,5,6&28&1\\
3&1,4,5,6&2&1\\
3&2,3,4,5&0&1\\
3&2,3,4,6&0&1\\
3&2,3,5,6&0&1\\
3&2,4,5,6&0&1\\
3&3,4,5,6&2&1\\
\hline
4&1,2,3,4,5&0&0\\
4&1,2,3,5,6&0&0\\
4&1,2,3,4,6&0&0\\
4&1,2,4,5,6&0&0\\
4&1,3,4,5,6&1&0\\
4&2,3,4,5,6&0&0\\
\hline
\end{tabular}
\end{center}

\end{appendix}
\newpage
\bibliographystyle{JHEP}
\bibliography{refs}
\end{document}